\documentclass[epsfig,graphics,twocolumn,floatfix,a4paper,showpacs]{revtex4}
\usepackage{amsmath,amsfonts,amssymb,graphicx,color,times,bbm,psfrag,dsfont,hyperref}
\usepackage[english]{babel}
\usepackage[latin1]{inputenc}
\usepackage[T1]{fontenc}

\newif\ifhyper
\hypertrue
\ifhyper
\hypersetup{
   citecolor = {green},
   colorlinks = {true}, 
   urlcolor = {blue} 
}
\fi

\newcommand{\beq}{\begin{equation}}
\newcommand{\eeq}{\end{equation}}
\newcommand{\beqa}{\begin{eqnarray}}
\newcommand{\eeqa}{\end{eqnarray}}
\newcommand{\ket} [1] {\vert #1 \rangle}
\newcommand{\bra} [1] {\langle #1 \vert}
\newcommand{\braket}[2]{\langle #1 | #2 \rangle}

\usepackage{graphicx}
\usepackage[mathscr]{eucal}
\usepackage{amsmath}
\usepackage{amssymb}
\usepackage{epstopdf}
\usepackage{epsfig}
\def\one{\ensuremath{\hbox{$\mathrm I$\kern-.6em$\mathrm 1$}}}

\begin{document} 


\title{Classical simulation of infinite-size quantum lattice systems in two spatial dimensions}
\author{J. Jordan$^1$, R. Or\'us$^1$, G. Vidal$^1$, F. Verstraete$^2$, J. I. Cirac$^3$}

\affiliation{$^1$School of Physical Sciences, The University of Queensland, QLD 4072, Australia \\
$^2$ Fakult${\it \ddot{{\rm {\it  a}}}}$t f${\it \ddot{{\rm {\it u}}}}$r Physik, Universit${\it \ddot{{\rm {\it  a}}}}$t Wien, Boltzmanngasse 3, A-1090 Wien \\
$^3$ Max-Planck-Institut f${\it \ddot{{\rm {\it u}}}}$r Quantenoptik, Hans Kopfermann-Str. 1, Garching, D-85748, Germany}

\begin{abstract}
 
We present an algorithm to simulate two-dimensional quantum lattice systems in the thermodynamic limit. Our approach builds on the {\em projected entangled-pair state} algorithm for finite lattice systems [F. Verstraete and J.I. Cirac, cond-mat/0407066] and the infinite {\em time-evolving block decimation} algorithm for infinite one-dimensional lattice systems [G. Vidal, Phys. Rev. Lett. 98, 070201 (2007)]. The present algorithm allows for the computation of the ground state and the simulation of time evolution in infinite two-dimensional systems that are invariant under translations. We demonstrate its performance by obtaining the ground state of the quantum Ising model and analysing its second order quantum phase transition. 

\end{abstract}

\maketitle

Strongly interacting quantum many-body systems are of central importance in several areas of science and technology, including condensed matter and high-energy physics, quantum chemistry, quantum computation and nanotechnology. From a theoretical perspective, the study of such systems often poses a great computational challenge. Despite the existence of well-stablished numerical techniques, such as exact diagonalization, quantum monte carlo \cite{QMC}, the density matrix renormalization group \cite{DMRG} or series expansion \cite{SE} to mention some, a large class of two-dimensional lattice models involving frustrated spins or fermions remain unsolved.

Fresh ideas from quantum information have recently led to a series of new simulation algorithms based on an efficient representation of the lattice many-body wave-function through a \emph{tensor network}. This is a network made of small tensors interconnected according to a pattern that reproduces the structure of entanglement in the system. Thus, a \emph{matrix product state} (MPS) \cite{MPS}, a tensor network already implicit in the density matrix renormalization group, is used in the time-evolving block decimation (TEBD) algorithm to simulate time evolution in one-dimensional lattice systems \cite{TEBD}, whereas a \emph{tensor product state} \cite{TPS} or \emph{projected entangled-pair state} (PEPS) \cite{PEPS} is the basis to simulate two-dimensional lattice systems. In turn, the \emph{multi-scale entanglement renormalization ansatz} accuratedly describes critical and topologically ordered systems \cite{MERA}.

In this work we explain how to modify the PEPS algorithm of Ref. \cite{PEPS} to simulate two-dimensional lattice systems in the thermodynamic limit. By addressing an infinite system directly, the infinite PEPS (iPEPS) algorithm can analyse bulk properties without dealing with boundary effects or finite-size corrections. This is achieved by generalizing, to two dimensions, the basic ideas underlying the infinite TEBD (iTEBD) \cite{iTEBD}. Namely, we exploit translational invariance ($i$) to obtain a very compact PEPS description with only two independent tensors and ($ii$) to simulate time evolution by just updating these two tensors. We describe the essential new ingredients of the iPEPS algorithm, which is based on numerically solving a transfer matrix problem with an MPS. We then use it to compute the ground state of the quantum Ising model with transverse magnetic field, evaluate local observables, identify the critical point of its second order quantum phase transition and estimate the critical exponent $\beta$.

We point out that the algorithms of Ref. \cite{TPS} have already addressed the computation of the ground state of infinite two-dimensional systems, by analysing an infinite transfer matrix problem with a MPS. A major difference in our approach is how this is handled. Instead of DMRG techniques (which consider an increasingly large chain with a finite MPS), we use the iTEBD algorithm \cite{iTEBD}, based on a power method that uses an infinite MPS (iMPS) from the start. This seems to significantly improve the results reported in Ref. \cite{TPS}.

\begin{figure}[h]
\includegraphics[width=8.5cm]{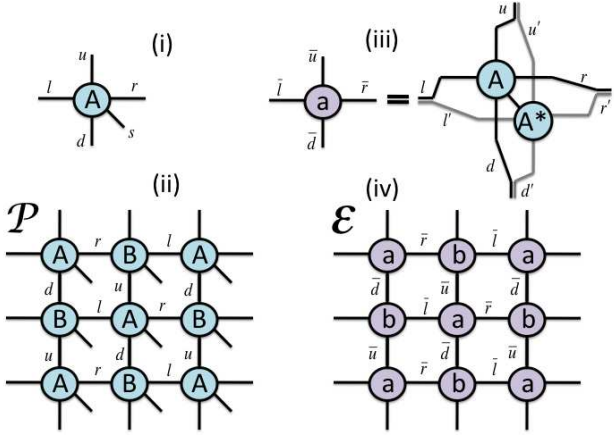}
\caption{(color online) Diagramatic representations of ($i$) a PEPS tensor $A_{sudlr}$ with one physical index $s$ and four inner indices $u$,$d$, $l$ and $r$; ($ii$) local detail of the tensor network $\mathcal{P}$ for an iPEPS. Copies of tensors $A$ and $B$ are connected through four types of links; ($iii$) reduced tensor $a$ of Eq. (\ref{eq:reduced}); and ($iv$) local detail of the tensor network $\mathcal{E}$.}
\label{fig:PEPS}
\end{figure}

\textbf{Finite PEPS algorithm.---} We start by recalling some basic facts of the PEPS algorithm for a finite system \cite{PEPS}. Consider a two-dimensional lattice $\mathcal{L}$ where each site, labeled by a vector $\vec{r}$, is represented by a Hilbert space $V^{[\vec{r}]} \cong \mathbb{C}^d$ of finite dimension $d$, so that the Hilbert space of the lattice is $V_{\mathcal L} = \bigotimes_{\vec{r}\in \mathcal{L}} V^{[\vec{r}]}$. For concreteness, we address the case of a square lattice, with $N\times N$ sites labelled by a pair of integers $\vec{r} = (x,y)$,  $~x,y=1,\cdots,N$. [However, the key ingredients of the algorithm for infinite systems to be considered here are still valid for any type of regular lattice.] The model is further characterized by a Hamiltonian $H = \sum_{\vec{r}_1,\vec{r}_2} h^{[\vec{r}_1\vec{r}_2]}$ that decomposes as a sum of terms $h^{[\vec{r}_1\vec{r}_2]}$ involving pairs of nearest neighbor sites $\vec{r}_1,\vec{r}_{2}\in \mathcal{L}$. A pure state $\ket{\Psi} \in V_{\mathcal{L}}$ of the lattice is represented by a PEPS, namely a set of $N \times N$ tensors $\{A^{[\vec{r}]}\}_{\vec{r}\in \mathcal{L}}$, interconnected into a network $\mathcal{P}$ that follows the pattern of $\mathcal{L}$ (Figs. \ref{fig:PEPS}.i and \ref{fig:PEPS}.ii). Tensor $A^{[\vec{r}]}_{sudlr}$ is made of complex numbers labelled by one {\em physical} index $s$ and four {\em bond} indices $u$, $d$, $l$ and $r$. The physical index runs over a basis $\{\ket{s}\}_{s=1,\cdots,d}$ of $V^{[\vec{r}]}$, whereas each bond index takes $D$ values and connects the tensor with the tensors in nearest neighbor sites. 
Thus, $\ket{\Psi}$ is written in terms of $O(N^2D^4d)$ parameters, from which the $d^N$ complex amplitudes $\braket{s_{(1,1)} s_{(1,2)} \cdots s_{(N,N)}}{\Psi}$ can be recovered by fixing the physical index of each tensor $A^{[\vec{r}]}$ in $\mathcal{P}$ and by contracting all the bond indices.

Given a PEPS for some state $\ket{\Psi_0}\in V_{\mathcal{L}}$, the algorithm of Ref. \cite{PEPS} allows to perform e.g. the following two tasks: ($i$) computation of expected values $\bra{\Psi_0}O\ket{\Psi_0}$ for a local operator $O$, such as the energy, a local order parameter or two-point correlators, and ($ii$) update of the PEPS after a gate $g^{[\vec{r}_1\vec{r}_2]}$ has been applied on two nearest sites $\vec{r}_1,\vec{r}_2\in \mathcal{L}$. The second task can be used to simulate an evolution according to Hamiltonian $H$, both in real time and in imaginary time,
\begin{equation}
	\ket{\Psi_t} = e^{-iHt}\ket{\Psi_0}, ~~~~~~ \ket{\Psi_{\tau}} = \frac{e^{-H\tau}\ket{\Psi_0}}{||e^{-H\tau}\ket{\Psi_0}||},
\label{eq:time_evolution}
\end{equation}
in the sense of obtaining a new PEPS representation that approximates the states $\ket{\Psi_t}$ and $\ket{\Psi_{\tau}}$. This is achieved by breaking the evolution operators $e^{-iHt}$ and $e^{-H\tau}$ into a sequence of local gates, using a Suzuki-Trotter expansion \cite{Trotter}, and by updating the PEPS after applying each of these gates. In particular, one can approximate the ground state of Hamiltonian $H$ through simulating an evolution in imaginary time for large time $\tau$, starting from a product state $\ket{\Psi_0}$ (for which a PEPS can be trivially constructed). 

Let $\mathcal{E}$ denote the network made by the $N\times N$ reduced tensors $a^{[\vec{r}]}$ (Figs. \ref{fig:PEPS}.iii and \ref{fig:PEPS}.vi),
\begin{equation}
a^{[\vec{r}]}_{\bar{u}\bar{d}\bar{l}\bar{r}} \equiv \sum_{s} A^{[\vec{r}]}_{s~udlr} (A^{[\vec{r}]}_{s~u'd'l'r'})^*,
\label{eq:reduced}
\end{equation}
where $\bar{u}$ represents the double bond index $(u,u')$ and the physical index $s$ has been contracted, and let $\vec{r}_1,\vec{r}_2\in\mathcal{L}$ denote two nearest neighbor sites.
Then the \emph{environment} $\mathcal{E}^{[\vec{r}_1, \vec{r}_2]}$ for these two sites is the network obtained by removing the tensors $a^{[\vec{r}_1]}$ and $a^{[\vec{r}_2]}$ from $\mathcal{E}$. By "\emph{contracting a tensor network}" we mean "\emph{summing over all the indices that connect any two tensors of the network}". It turns out that both ($i$) the computation of an expected value $\bra{\Psi}O^{[\vec{r}_1,\vec{r}_2]}\ket{\Psi}$ and ($ii$) the update of the PEPS after a gate $g^{[\vec{r}_1\vec{r}_2]}$ can be achieved by contracting $\mathcal{E}^{[\vec{r}_1, \vec{r}_2]}$. 
However, the cost of this contraction grows exponentially with $N$. The core of the PEPS algorithm \cite{PEPS} is an approximate, \emph{efficient} (quadratic in $N$) scheme to contract $\mathcal{E}^{[\vec{r}_1, \vec{r}_2]}$, based on MPS  simulation techniques.

\begin{figure}
\includegraphics[width=8.5cm]{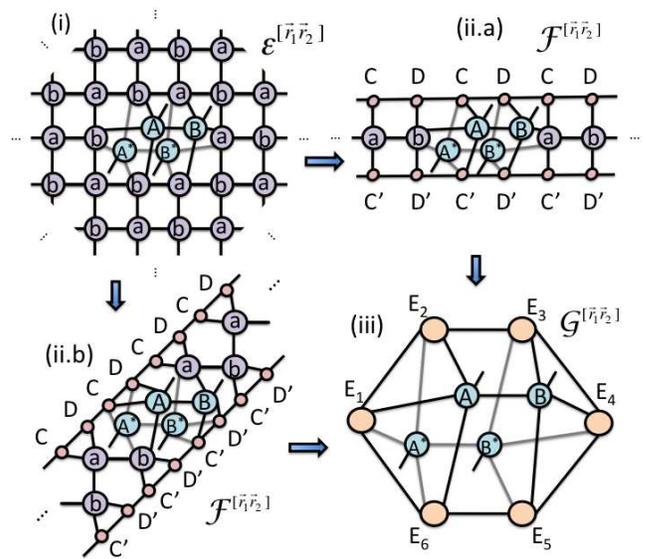}
\caption{The environment $\mathcal{E}^{[\vec{r}_1, \vec{r}_2]}$ for a link of type $r$ is first approximated by an infinite strip $\mathcal{F}^{[\vec{r}_1, \vec{r}_2]}$ and then by a six-tensor network $\mathcal{G}^{[\vec{r}_1, \vec{r}_2]}$. These reductions can be performed according to either a vertical/horizontal scheme ($ii.a$) or a diagonal scheme ($ii.b$). Tensors $A,A^\star,B$ and $B^\star$ are not part of the environment.}
\label{fig:Reductions}
\end{figure}

\begin{figure}
\includegraphics[width=8.0cm]{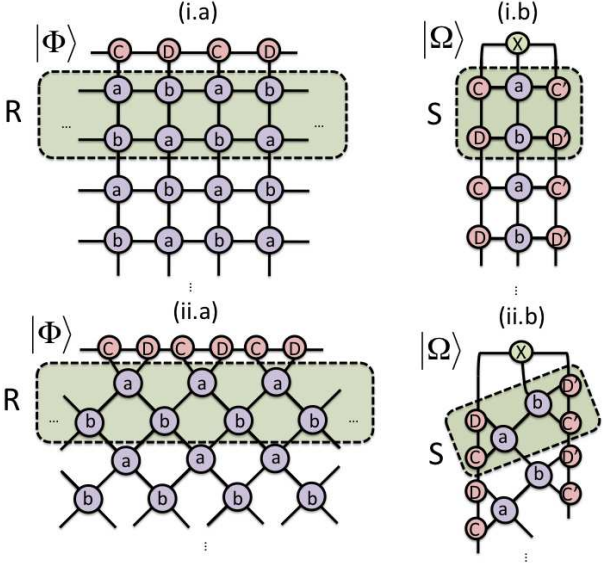}
\caption{Transfer matrices $R$ (i.a) and $S$ (i.b) for the vertical/horizontal contraction scheme. Multiplication of an iMPS by $R$ using the iTEBD algorithm comes at a computational time that scales as $O(\chi^3D^6+\chi^2D^8d)$ (similar costs apply to multiplying by $S$). This cost is lower in diagonal contraction scheme (ii.a) and (ii.b), namely $O(\chi^3D^4+\chi^2D^6d)$, but a slightly larger $\chi$ is required in order to retain the same accuracy.}
\label{fig:Transfer}
\end{figure}

\textbf{Infinite PEPS algorithm.---} In order to consider the limit of an infinite lattice, $N\rightarrow \infty$, where both $\ket{\Psi}$ and $H$ are invariant under shifts by one lattice site, we need to understand how to efficienlty contract an infinite environment $\mathcal{E}^{[\vec{r}_1, \vec{r}_2]}$. Translational invariance allows us to represent the state $\ket{\Psi}$ in terms of only two tensors $A$ and $B$ that are repeated indefinitely in $\mathcal{P}$ (Fig. \ref{fig:PEPS}),
\begin{eqnarray}
	A^{[(x,x+2y)]} = A, ~~~A^{[(x,x+2y+1)]} = B, ~~x,y \in \mathbb{Z},
\end{eqnarray}
so that the iPEPS depends on just $O(D^4d)$ coefficients. Notice that $\mathcal{E}^{ [\vec{r}_1, \vec{r}_2]}$ is also made of infinitely many copies of just two reduced tensors $a$ and $b$, defined in terms of $A$ and $B$ according to Eq. (\ref{eq:reduced}). Then its contraction is achieved in two steps, as illustrated in Fig. (\ref{fig:Reductions}): first we approximate $\mathcal{E}^{[\vec{r}_1, \vec{r}_2]}$ with an infinite strip $\mathcal{F}^{[\vec{r}_1, \vec{r}_2]}$; then we approximate $\mathcal{F}^{[\vec{r}_1, \vec{r}_2]}$ with a small set of tensors $\mathcal{G}^{[\vec{r}_1, \vec{r}_2]} = \{G_1, \cdots, G_6\}$. This can be achieved using a \emph{vertical/horizontal} scheme or a \emph{diagonal} scheme as illustrated in Figs. \ref{fig:Reductions}-\ref{fig:Transfer}.

The first step considers a transfer matrix $R$ consisting of an infinite strip of reduced tensors $a$ and $b$ (Fig. \ref{fig:Transfer}). $R$ can be regarded as a linear operator acting on an infinite chain where each site is described by a vector space of dimension $D^2$. Let $\ket{\Phi}$ denote the \emph{dominant} eigenvector of $R$---that is, the eigenvector of $R$, $R\ket{\Phi} = \lambda\ket{\Phi}$, with the eigenvalue $\lambda$ of largest absolute value. Here we assume that the dominant eigenvector is unique \cite{unique}. By construction $R$ is invariant under shifts by two sites of the infinite chain -- and so is $\ket{\Phi}$. We use an iMPS, characterized by just two tensors $\{C,D\}$ and with Schmidt rank $\chi$, to represent an approximation of $\ket{\Phi}$. We obtain this iMPS by simulating (repeated) multiplication by $R$ on an initial state $\ket{\Phi_0}$ with the iTEBD algorithm \cite{iTEBD} and by using the fact that
\begin{equation}
 \ket{\Phi} = \lim_{p\rightarrow \infty} \frac{R^p\ket{\Phi_0}}{||R^p\ket{\Phi_0}||}.
\end{equation}
The iMPS for $\ket{\Phi}$ accounts for an infinite half plane of the environment $\mathcal{E}^{[\vec{r}_1, \vec{r}_2]}$. Similarly, we use another iMPS with tensors $\{C',D'\}$ to encode the left dominant eigenvector $\bra{\Phi'}$ of $R$, $\bra{\Phi'}R = \lambda \bra{\Phi'}$, which also accounts for an infinite half plane. Then $\mathcal{F}^{[\vec{r}_1, \vec{r}_2]}$ is built from these two iMPS and a strip of reduced tensors $a$ and $b$. 

In the second step, a transfer matrix $S$ is defined in terms of the tensors $\{a,b,C,D,C',D'\}$ (Fig. \ref{fig:Transfer}). $S$ can be regarded as a linear operator acting on three sites with local vector space dimensions $\chi$, $D^2$ and $\chi$. Again, its dominant eigenvector $\ket{\Omega}$, encoded in a three-legged tensor $X$, is computed from an initial state $\ket{\Omega_0}$ using the fact that
\begin{equation}
 \ket{\Omega} = \lim_{q\rightarrow \infty} \frac{S^q\ket{\Omega_0}}{||S^q\ket{\Omega_0}||}.
\end{equation}
Let $X'$ be the tensor for the left dominant eigenvector $\bra{\Omega'}$ of $S$. Then $\mathcal{G}^{[\vec{r}_1 \vec{r}_2]}$ is a (circular) MPS consisting of the six tensors $\{C,D,C',D',X,X'\}$.

{\bf Simulation of time evolution.---}
We decompose the Hamiltonian as $H=H_r + H_d + H_l + H_u$, where the operator $H_r = \sum_{(\vec{r}_1,\vec{r}_2)_r} h^{[\vec{r}_1\vec{r}_2]}$ collects all interactions along $r$-links (and similarly for $d$-, $l$- and $u$-links), and consider a Suzuki-Trotter expansion of the time-evolution operator $e^{-iHt}$ of Eq. (\ref{eq:time_evolution}) in terms of operators $e^{-i H_r\delta t}$, $e^{-i H_d\delta t}$, $e^{-i H_l\delta t}$ and $e^{-iH_u\delta t}$, where $\delta t$ is some small time step. 
Each of these operators further decomposes into a product of identical two-site unitary gates $g \equiv e^{-ih\delta t}$ acting on all pairs of sites connected by a link of the proper type. For instance, for links of type $r$ we have
\begin{equation}
	e^{-iH_r\delta t} = \prod_{(\vec{r},\vec{r}')_r} g^{[\vec{r}\vec{r}']}.
\label{eq:Ur}
\end{equation}
Without loss of generality, we need to address only the update of tensors $A$ and $B$ after applying $e^{-iH_r\delta t}$
to $\ket{\Psi}$. Let us assume that the gate $g$ is applied on just \emph{one} of the $r$-links. In that case, in order to update the iPEPS we would ($i$) compute the environment for that specific $r$-link following Figs. \ref{fig:Reductions}-\ref{fig:Transfer}, and ($ii$) determine the optimal new tensors $A'$ and $B'$ for the link, using the optimization techniques of \cite{PEPS} (sweeping over just the two sites involved). We notice, however, that the above $A'$ and $B'$ already define an iPEPS for $e^{-iH_r\delta}\ket{\Psi}$ -- that is, with gates $g$ acting on \emph{all} $r$-links. Indeed, this follows from translation invariance and the fact that a {\em unitary} gate $g$ on a given $r$-link does not affect the environment on any other $r$-link. In other words, the update of tensors $A$ and $B$ on each $r$-link is identical and need only be performed once. 

The above argumentation fails for an evolution $e^{-H\tau}$ in imaginary time, since the gate $g'\equiv e^{-h\delta\tau}$ is no longer unitary. In this case, a gate applied on, say, an $r$-link modifies the environment on the rest of $r$-links. Nevertheless, as in one-dimensional systems \cite{iTEBD}, the same algorithm can still be used to find the ground state of the system through imaginary-time evolution, provided that a sufficiently small $\delta \tau$ (leading to almost unperturbed environments) is used at the last stages of the simulation.

\begin{figure}
\includegraphics[width=8.5cm]{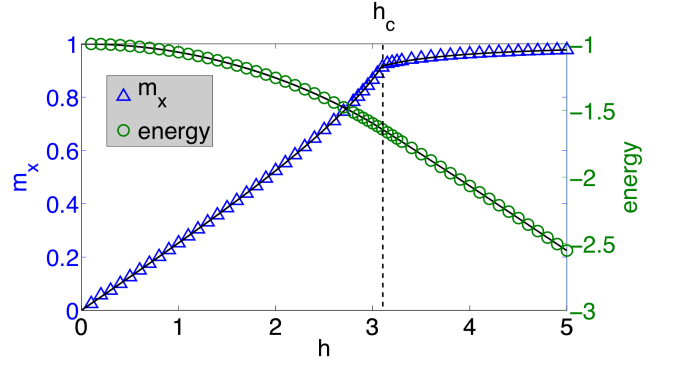}
\caption{Transverse magnetization $m_x$ and energy per site $e$ as a function of the transverse magnetic field $h$. The continuous line shows series expansion results (to 26th and 16th order in perturbation theory) for $h$ smaller and larger than $h_c\approx 3.044$ \cite{SEresults}. Increasing $D$ leads to a lower energy per site $e$. For instance, at $h=3.1$, $e(D=2) \approx -1.6417$ and $e(D=3) \approx -1.6423$.}
\label{fig:mag_energy}
\end{figure}

\begin{figure}
\includegraphics[width=8.5cm]{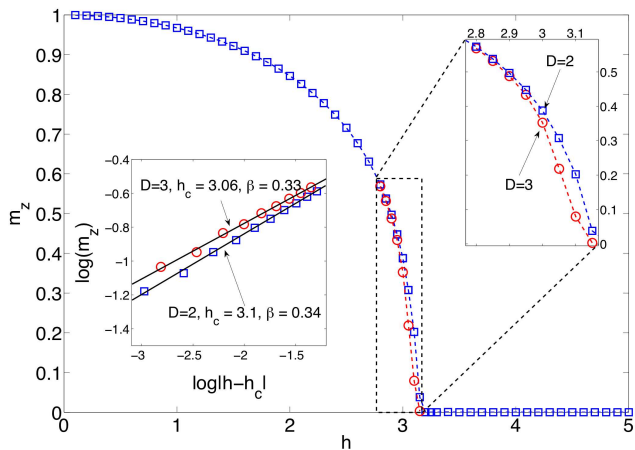}
\caption{(color online) Magnetization $m_z(\lambda)$ as a function of the transverse magnetic field $\lambda$. Dashed lines are a guide tp the eye. We have used the diagonal scheme for $(D,\chi) = (2,20)$, $(3,25)$ and $(4,35)$ \cite{chi} (the vertical/horizontal scheme leads to comparable results with slightly smaller $\chi$.) The inset shows a log plot of $m_z$ versus $|\lambda - \lambda_c|$, including our estimate of $\lambda_c$ and $\beta$. The continuous line shows the linear fit.
}
\label{fig:order}
\end{figure}

\begin{figure}
\includegraphics[width=8.5cm]{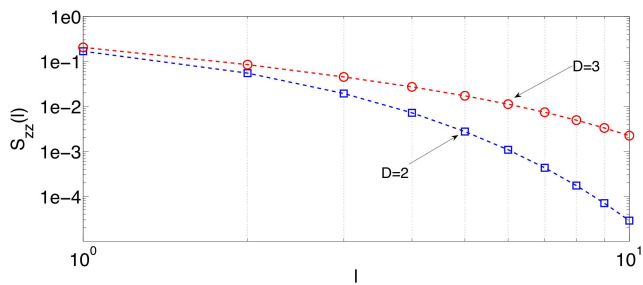}
\caption{(color online) Two-point correlator $S_{zz}(l)$ near the critical point, $\lambda = 3.05$. For nearest neighbors, the correlator quickly converges as a function of $D$, whereas for long distances we expect to see convergence for larger values of $D$.
}
\label{fig:corr}
\end{figure}

{\bf Quantum phase transition.---} To demonstrate the performance of the iPEPS algorithm, we have simulated an evolution in imaginary time to obtain the ground state $\ket{\Psi_{\lambda}}$ of the quantum Ising model with transverse magnetic field,
\begin{equation}
	H(\lambda) \equiv - \sum_{(\vec{r},\vec{r}')} \sigma_z^{[\vec{r}]}\sigma_z^{[\vec{r}']} -  \lambda\sum_{\vec{r}}\sigma_x^{[\vec{r}]}.
\label{eq:qIsing}
\end{equation}
Then we have computed the energy per site $e$ and the transverse and parallel magnetizations $m_x$ and $m_z$ (Figs. \ref{fig:mag_energy}-\ref{fig:order}),
\begin{equation}
	m_x(\lambda) = \bra{\Psi_{\lambda}} \sigma_x \ket{\Psi_{\lambda}}, ~~~~~	m_z(\lambda) = \bra{\Psi_{\lambda}} \sigma_z \ket{\Psi_{\lambda}}, 
\end{equation}
and the two point correlator $S_{zz}(l)$ (Fig. \ref{fig:corr})
\begin{equation}
	S_{zz}(l) \equiv \bra{\Psi_{\lambda}} \sigma_z^{[\vec{r}]}\sigma_z^{[\vec{r}+l\hat{e}_x]} \ket{\Psi_{\lambda}} - (m_{z})^2.
\end{equation}

\begin{table}
	\begin{tabular}{|c |c| c |c |c |c|}
	\hline
	   & QMC {\scriptsize Ref. \cite{QMCresults} } &  D=2 ~{\scriptsize iPEPS} & D=3 ~{\scriptsize iPEPS} &  D=3~{\scriptsize VDMA Ref. \cite{TPS}}\\
	\hline\hline
	$\lambda_c$ & 3.044 & 3.10 & 3.06 & 3.2 \\
	\hline
	$\beta$ & 0.327 &	0.346 & 0.332 & -- \\
	\hline 
	\end{tabular}
	\caption{Critical point and exponent $\beta$ as a function of D.}
	\label{tab:Critical}
\end{table}

Comparison with series expansion results of Ref. \cite{SEresults} shows remarkable agreement for all local observables on both sides of the critical point, which Monte Carlo calculations \cite{QMCresults} indicate to be at magnetic field $\lambda_{MC}\approx 3.044$.
We also obtain accurate estimates of the critical magnetic field $\lambda_c$ and critical exponent $\beta$, which for  $D=2$ and $D=3$ agree with Monte Carlo results within $5.8 \%$ and $1.5\%$ respectively. 

In conclusion, we have presented an algorithm to simulate infinite two-dimensional lattice systems. We have tested its performance in the context of the quantum Ising model, where our results can compete quantitatively with those obtained using long-established methods, such as quantum Monte Carlo \cite{QMC} or perturbative series expansions \cite{SE}. The iPEPS algorithm can now be applied to address problems beyond the reach of quantum Monte Carlo (since it has no {\em sign problem}) and series expansion methods (since it does not rely on an expansion around an exactly solvable model). Thus we expect it to become a useful new tool in the study of strongly interacting lattice models.

Support from the Australian Research Council (APA, DP0878830 and FF0668731) is acknowledged.


\begin{thebibliography}{100}
\bibitem{unique} We find that the dominant eigenvector is unique both at the disordered and ordered phases of the quantum Ising model. In the ordered phase, spontaneous symmetry breaking seems to be directly responsible for this uniqueness.
\bibitem{chi} Given a value of the bond dimension $D$, we choose $\chi$ large enough so that the results do no longer depend significantly on $\chi$.
\end{thebibliography}
\end{document}